\documentclass[%
 aip,
 jap,
 amsmath,amssymb,
 preprint,%
]{revtex4-2}
\usepackage{xcolor}
\usepackage{mathrsfs}
\usepackage{graphicx}%
\usepackage{dcolumn}%
\usepackage{bm}%
\usepackage{siunitx}
\usepackage{makecell}
\usepackage[mathlines]{lineno}%
\relax %

\begin{document}

\preprint{AIP/123-QED}

\title{First-Principles Prediction of Nonlinear Optical Response in TiO$_2$ for High-Power Dielectric Mirror Applications}

\author{Koya Shimaoka}
\affiliation{Department of Electrical and Electronic Engineering, Graduate School of Engineering, Kobe University, Nada, Kobe 657-8501, Japan}

\author{Yusuke Kondo}
\affiliation{Osaka Research Institute of Industrial Science and Technology (ORIST), 2-7-1 Ayumino, Izumi, Osaka 594-1157, Japan}

\author{Kazunori Shibata}
\affiliation{Institute of Laser Engineering (ILE), The University of Osaka, 2-6 Yamadaoka, Suita, Osaka 565-0871, Japan}

\author{Mitsuharu Uemoto}
\email{uemoto@eedept.kobe-u.ac.jp}
\affiliation{Department of Electrical and Electronic Engineering, Graduate School of Engineering, Kobe University, Nada, Kobe 657-8501, Japan}

\begin{abstract}
Dielectric multilayer mirrors are essential components in optical experiments using high-power lasers, where titanium dioxide (TiO$_2$) is widely employed as a high-refractive-index dielectric material.
In this study, we investigate the nonlinear optical response of TiO$_2$ under intense ultrashort laser pulses using real-time first-principles electron-dynamics simulations based on time-dependent density functional theory (TDDFT).
We reveal intensity-dependent absorption driven by multiphoton excitation and optically excited free carriers, and simulate the resulting electron--light coupled dynamics in TiO$_2$ nanofilms using a multiscale Maxwell--TDDFT framework.
Direct evaluation of the reflected and transmitted fields shows reduced reflectance at high intensities ($I \sim 10^{13}~\mathrm{W}/\mathrm{cm}^2$), demonstrating a pronounced nonlinear optical response.
Furthermore, the optical response properties are compared among various stable and metastable crystalline phases of TiO$_2$---rutile, anatase, brookite, TiO$_2$-II, and TiO$_2$-B---as well as an amorphous supercell model.
These results provide microscopic insight into intensity-dependent optical degradation in TiO$_2$-based dielectric optical components exposed to intense femtosecond laser fields.
\end{abstract}

\maketitle

\section{Introduction}

High-power lasers are widely used across diverse fields, from industrial applications to fundamental physics research, including laser machining \cite{stoian2020advances, miyaji2025stable, kinouchi2023laser}, extreme ultraviolet generation \cite{versolato2019physics}, particle acceleration, and laser-based active debris removal technology \cite{monroe1993space, SHIBATA2024281, khomich2025laser}.
Peak intensities achievable with current laser technology have increased rapidly \cite{gales2018extreme}. For instance, the world-record intensity of $1.1 \times 10^{23}~\mathrm{W}/\mathrm{cm}^2$ was reported for the CoReLS petawatt laser \cite{yoon2021realization} in 2021.
Furthermore, extreme-intensity lasers are enabling the exploration of new frontiers in physics, including experiments to observe vacuum quantum electrodynamics (QED) effects near the Schwinger limit \cite{PhysRev.82.664, RRP-S145, PhysRevD.90.092003, Della_Valle_2013, PhysRevA.109.063503, Shibata2022EPJD, Shibata_2022, PhysRevA.104.063513, Shibata2021EPJD, Shibata2020, Shibata_2019}.
There are demands for optical materials capable of withstanding high-power laser irradiation.

Dielectric multilayer mirrors, composed of alternating layers of two dielectric materials with differing refractive indices, are widely utilized in laser optical systems. Their popularity stems from their ability to achieve high reflectance and minimal absorption owing to reduced Ohmic losses \cite{laurence2020mirrors}.
However, their applicability is limited in the extreme-intensity regime due to strong-field nonlinear processes such as multiphoton absorption and ablation damage \cite{kozlov2019mechanisms, laurence2017role}, which are amplified by local field enhancement originating from microscopic structural defects.
The development of high-power laser mirrors is a significant engineering challenge that requires not only advanced fabrication techniques but also simulation-driven structural design that accounts for nonlinear interactions accurately.

Computational electromagnetic simulations are indispensable tools for designing photonic materials, including dielectric multilayers.
Numerical solvers for Maxwell's equations, such as the finite-difference time-domain (FDTD) method \cite{yee1966numerical}, have been used extensively.
However, accurate modeling of nonlinear responses requires a detailed description of quantum-mechanical electron dynamics.
Time-dependent density functional theory (TDDFT) \cite{runge1984density} is frequently employed in theoretical and computational studies to predict optical responses in the high-intensity regime.
Previously, we implemented a coupled Maxwell--TDDFT multiscale framework that integrates FDTD-based electromagnetics with TDDFT \cite{yabana2012time}; this framework enables the simulation of interactions between nanomaterials and high-intensity lasers while accounting for nonperturbative nonlinear optical responses \cite{lee2014first, sato2015time, yamada2024interaction, uemoto2021first, uemoto2022first}.

In this paper, we apply these computational approaches to investigate the optical properties of TiO$_2$, which is a high-refractive-index material commonly employed in such structures.
Specifically, we perform the following first-principles simulations:
(1) we analyze electron dynamics in the rutile phase bulk TiO$_2$ under strong laser fields using TDDFT to establish fundamental nonlinear interactions; (2) we estimate the ablation threshold by comparing the absorbed energy with critical energetic thresholds, such as melting, cohesive, and bond-breaking energies; (3) we simulate light propagation using a multiscale Maxwell--TDDFT framework to evaluate reflectance and absorption coefficients for a TiO$_2$ thin film, thereby elucidating the influence of nonlinear optical effects on reflection characteristics.
Furthermore, for a realistic description of industrially fabricated amorphous films, (4) we extend our analysis to various stable and metastable crystalline polymorphs \cite{diebold2003surface} (e.g., rutile, anatase, brookite, $\mathrm{TiO}_2\mathrm{-II}$, and $\mathrm{TiO}_2\mathrm{-B}$) as well as amorphous supercell models generated via molecular dynamics simulations \cite{ batatia2025foundation}.
Our results provide microscopic insight into degradation in TiO$_2$-based dielectric optical components exposed to intense femtosecond laser fields and thereby contribute to the establishment of a theoretical framework for the design of novel materials in high-power laser applications.

\section{Methods}

We employ real-time first-principles electron-dynamics calculations based on TDDFT.
The time evolution of the Bloch orbital $u_{b \bm{k}}(\bm{r}; t)$ under an external vector potential $\bm{A}(t)$ is governed by the time-dependent Kohn--Sham (TDKS) equation \cite{runge1984density}:
\begin{align}
  \mathrm{i} \hbar \frac{\partial}{\partial t} u_{b \bm{k}}(\bm{r}; t)
  &=
  \left[
    \frac{1}{2m}
    \left(
      -\mathrm{i}\hbar\nabla + \hbar\bm{k} + \frac{e}{c}\bm{A}(t)
    \right)^{2}
    + \hat{v}_{\mathrm{ion}} + v_{\mathrm{H}} + v_{\mathrm{XC}}
  \right]
  u_{b \bm{k}}(\bm{r}; t)
  \;,
\end{align}
where $\hat{v}_{\mathrm{ion}}$, $v_{\mathrm{H}}$, and $v_{\mathrm{XC}}$ represent the electron--ion interaction (pseudopotential), the electron--electron interaction (Hartree potential), and the exchange-correlation potential, respectively.
Throughout this paper, electromagnetic quantities are expressed in the Gaussian unit system.
The macroscopic averaged current density $\bm{j}(t)$ is obtained from the expectation value of the velocity operator, summed over occupied bands $b$ and Brillouin-zone sampling points $\bm{k}$ (with cell volume $V$):
\begin{align}
  \bm{j}(t)
  =&
  \frac{1}{V}
  \sum_{b \bm{k}}
  \frac{-2e}{m}
  \left[
    \mathrm{Re}~
    \iiint
    u^\ast_{b \bm{k}}(\bm{r}; t)
    \left(
      \frac{\hbar}{\mathrm{i}}  \nabla + \frac{\bm{A}(t)}{c}
    \right)
    u_{b \bm{k}}(\bm{r}; t)
    \;
    \mathrm{d}\bm{r}
  \right]
  +
  \bm{j}_\mathrm{pseudo}(t)
  \;,
\end{align}
where the term $\bm{j}_{\mathrm{pseudo}}$ accounts for the nonlocal pseudopotential contribution.

To describe the propagation of laser pulses within the material, we employ a Maxwell--TDDFT multiscale method that utilizes two distinct spatial scales: a macroscopic grid for electromagnetic fields and a microscopic grid for electronic dynamics.
On the optical wavelength scale, the time evolution of electromagnetic fields is obtained by solving Maxwell's equations with the FDTD method on a coarse macroscopic grid $\bm{R}$:
\begin{align}
 \nabla \times \left[\nabla \times \bm{A}_{\bm{R}}(t)\right]
 +
 \frac{1}{c^2}
 \frac{\partial^2}{\partial t^2}
  \bm{A}_{\bm{R}}(t)
 =
 -\frac{4\pi}{c}
  \bm{j}_{\bm{R}}(t).
\end{align}
Each macroscopic point $\bm{R}$ is coupled to an independent periodic solid-state system described by $u_{b\bm{k}\bm{R}}$, in which electronic dynamics are computed via TDDFT simulations.
This coupling provides the constitutive response for the electromagnetic fields, enabling the simulation of interactions between nanomaterials and high-intensity lasers while accounting for nonperturbative nonlinear optical responses.
The detailed explanation is given in previous works \cite{yabana2012time, uemoto2021first, sato2015time}.

\section{Computation Model}

\begin{figure}[h!tbp]
  \centering
  \includegraphics[width=0.9\textwidth]{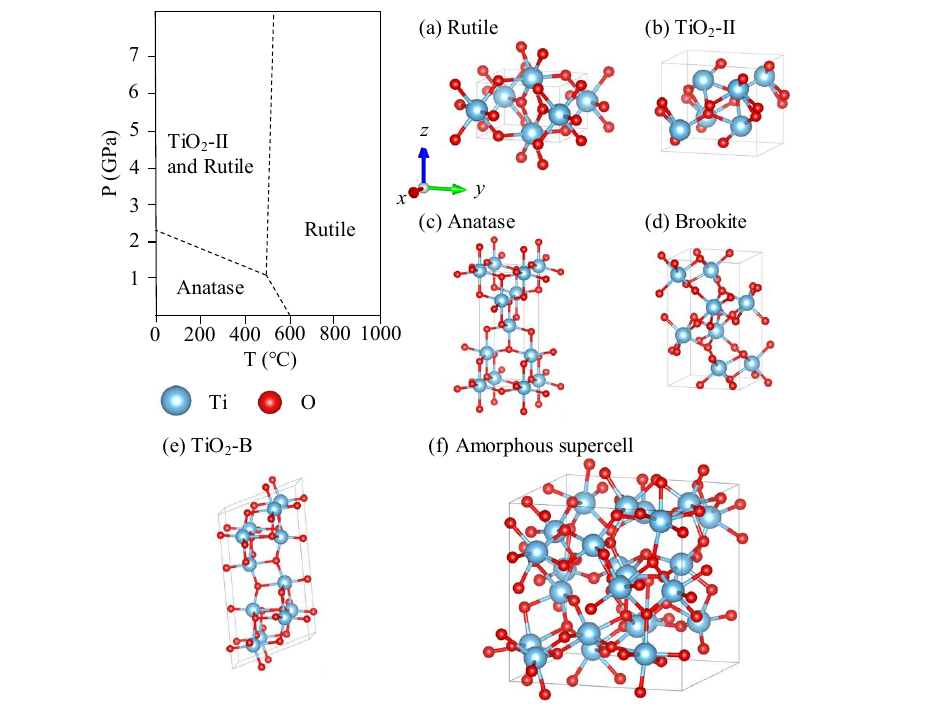}
  \caption{
    \label{fig:model}
    Bulk crystal structures of TiO$_2$:
    (a)~rutile phase,
    (b)~TiO$_2$-II,
    (c)~anatase,
    (d)~brookite,
    (e)~TiO$_2$-B, and
    (f)~amorphous supercell model.
    (The plotted phase diagram of TiO$_2$ is based on data extracted from Jamieson and Olinger~\cite{jamieson1969pressure} and Nie et al.~\cite{nie2009doping}.
    Figures of atomic structures are created using the visualization software VESTA \cite{momma2011vesta}.)
  }
\end{figure}

As shown in Fig.~\ref{fig:model}, solid TiO$_2$ exhibits several well-known crystalline phases, with forms such as rutile, anatase, and the TiO$_2$-II structure commonly observed under relatively low pressure and temperature \cite{jamieson1969pressure, nie2009doping}.
Previously, the nonlinear electron dynamics of anatase TiO$_2$ under strong laser fields have been studied in detail using real-time TDDFT \cite{sb2023ab}.
In this work, we first discuss the rutile-phase and its thin-film optical properties in detail, and then present a systematic comparison across multiple phases.

We employ a bulk rutile phase TiO$_2$ crystal (see Fig.~\ref{fig:model}(a)).
The structural model is based on a rectangular unit cell with dimensions of $4.5~\text{\AA} \times 4.5~\text{\AA} \times 2.8~\text{\AA}$ determined under equilibrium conditions.
We assume that the incident electromagnetic field is linearly polarized with a $\sin^2$ envelope pulse profile, where the pulse duration $\tau_\mathrm{pulse}$ is varied from $12~\mathrm{fs}$ to $50~\mathrm{fs}$.
The explicit expression for the pulse function is given in Appendix~\ref{sec:pulse_shape}.
The peak electric field amplitude $E_\mathrm{pulse}$ is varied from $0.027~\mathrm{V}/\text{\AA}$ to $2.74~\mathrm{V}/\text{\AA}$.
The central angular frequency $\omega_\mathrm{pulse}$ is chosen based on characteristic emission wavelengths of widely utilized solid-state lasers: namely, $\hbar\omega_\mathrm{pulse} = 1.55~\mathrm{eV}$ ($\lambda=800$~nm) for the Ti:sapphire laser and $\hbar\omega_\mathrm{pulse} = 1.17~\mathrm{eV}$ ($\lambda=1064$~nm) for the Nd:YAG laser \cite{boyd2008nonlinear}.

We perform our calculations using the SALMON (Scalable Ab-initio Light--Matter simulator for Optics and Nanoscience) package \cite{noda2019salmon}, which implements real-time TDDFT.
For the real-space implementation, the system is discretized on a spatial grid of $24 \times 24 \times 16$ points, and the Brillouin zone is sampled using a $4 \times 4 \times 6$ Monkhorst--Pack $k$-point grid.
The TDKS equation is integrated in time with a step size ranging from $3 \times 10^{-4}$~fs to $4 \times 10^{-4}$~fs by using time-reversal symmetric propagator \cite{marques2003octopus}.
We utilize the optimized norm-conserving Vanderbilt pseudopotentials (ONCVPSP) \cite{hamann2013optimized} provided by the PseudoDojo project \cite{van2018pseudodojo}, and employ the exchange-correlation functional treated within the local density approximation of Perdew and Zunger (PZ-LDA) \cite{perdew1981self}.
Furthermore, we extend our analysis to other stable or metastable crystalline phases as well as an amorphous supercell model (see Fig.~\ref{fig:model}(b)--(f)), the details of which are provided in Appendix~\ref{sec:param}.

\section{Results and Discussion}

\subsection{Ground-State Electronic System and Linear Responses}

\begin{figure*}[h!tbp]
    \centering
    \includegraphics[width=0.5\textwidth]{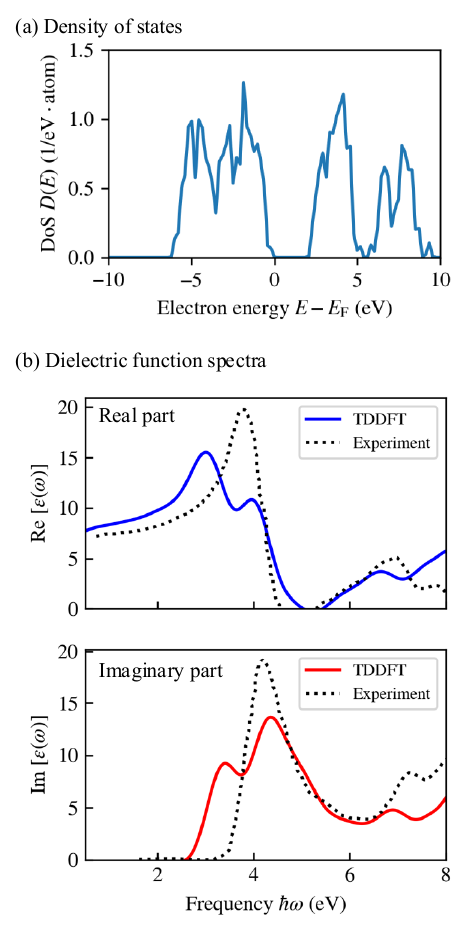}
    \caption{Density of states~(a) and complex dielectric function for $z$-polarization~(b) for rutile phase TiO$_2$ crystal.
Dotted lines correspond to experimental values reported in Ref.~\onlinecite{landmann2012electronic}.}
    \label{fig:epsilon}
\end{figure*}

First, we calculate the ground-state electronic structure and the dielectric function to characterize the linear optical response to weak electromagnetic fields.
As seen in Fig.~\ref{fig:epsilon}(a), the calculated bandgap energy $E_\mathrm{G}$ is $2.62~\mathrm{eV}$, which slightly underestimates the experimental value of $3~\mathrm{eV}$ due to the well-known tendency of the LDA to underestimate band gaps.
Despite this energy shift, the spectral features of the TDDFT dielectric functions shown in Fig.~\ref{fig:epsilon}(b) are qualitatively similar to those observed in previously reported experimental data \cite{landmann2012electronic}.
The imaginary part of $\epsilon$, which represents optical absorption, increases starting from $\hbar\omega \geq 2.62~\mathrm{eV}$, which corresponds to the absorption edge associated with $E_\mathrm{G}$.
Since the central frequency (photon energy) $\hbar \omega_\mathrm{pulse} = 1.55~\mathrm{eV}$ used in subsequent simulations is below this threshold, optical absorption is negligible.
(The anisotropy of dielectric functions for the other polarization direction is also presented in Supplementary Material S.~1.)

\subsection{Laser Pulse Response in Bulk Materials}
\label{sec:pulse}
\begin{figure}[h!tbp]
    \centering
    \includegraphics[width=\textwidth]{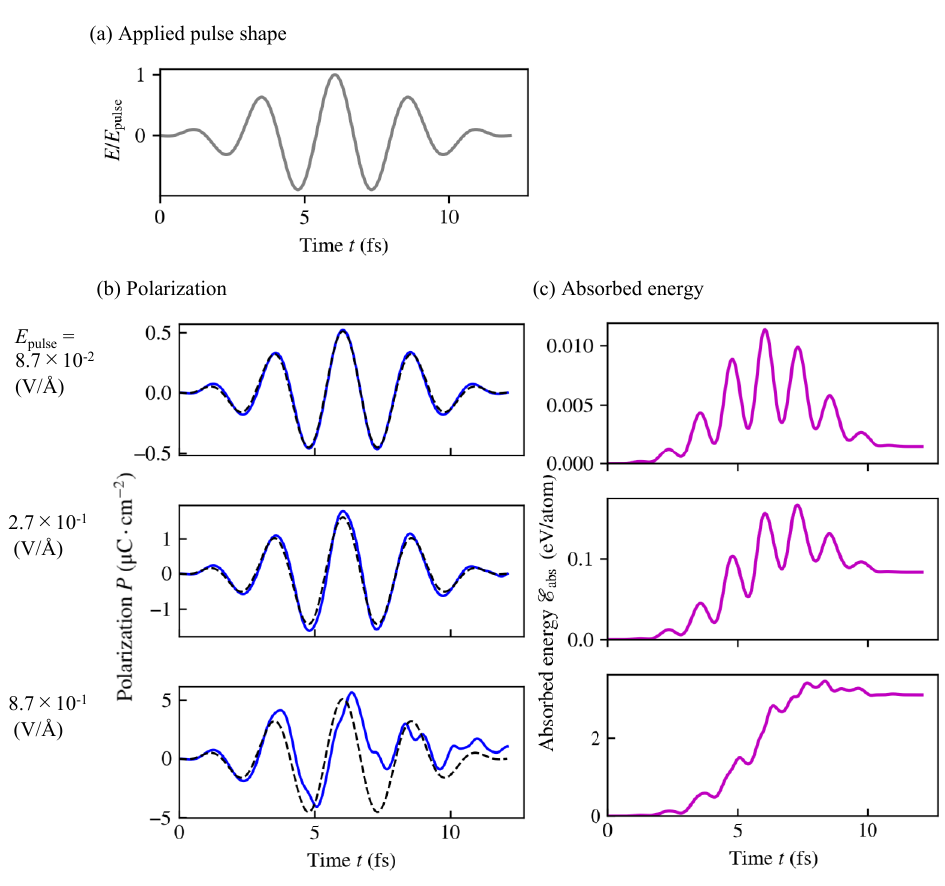}
    \caption{
Time-dependent dynamics of the electronic system under short and long pulse excitations:
(a)~Applied electric field amplitude,
(b)~induced polarization, and
(c)~absorbed energy calculated from Eq.~\eqref{eq:e_abs}.
The applied electric field amplitudes $E_\mathrm{pulse}$ range from $0.027~\mathrm{V}/\text{\AA}$ to $2.74~\mathrm{V}/\text{\AA}$.
   }
    \label{fig:pulse}
\end{figure}

To examine nonlinear optical properties, we investigate the real-time electronic response of bulk TiO$_2$ under intense laser irradiation; results for short and long pulse excitations are summarized in Fig.~\ref{fig:pulse}.
The induced polarization $\bm{p}(t)$ is defined as the time integral of the current density:
\begin{linenomath}
\begin{align}
\bm{p}(t)
=&
\int^t_0
 \bm{j}(t') \; \mathrm{d}t'.
\end{align}
\end{linenomath}
As illustrated in Fig.~\ref{fig:pulse}(b), the polarization
depends on laser intensity.
Under a weak field ($E_\mathrm{pulse}= 8.7\times 10^{-2}~\mathrm{V}/\text{\AA}$), the polarization waveform is nearly proportional to the applied field $E(t)$, as is typically observed in the linear response of insulating materials.
In contrast, under a strong field ($E_\mathrm{pulse}=2.7~\mathrm{V}/\text{\AA}$), a significant nonlinear response emerges during the latter half of the pulse, where the polarization acquires a phase shift relative to the driving field; this behavior indicates a metallic optical response \cite{yamada2024interaction}.
The absorbed energy shown in Fig.~\ref{fig:pulse}(c) provides a measure of the energy deposited into the electronic system during the pulse duration.
The absorbed energy is calculated based on Joule's law as follows:
\begin{linenomath}
\begin{align}
\mathscr{E}_\mathrm{abs}(t)
=&
\int^t_0
\bm{E}(t') \cdot \bm{j}(t') \; \mathrm{d}t',
\label{eq:e_abs}
\end{align}
\end{linenomath}
where the electric field is determined by $\bm{E}=-(1/c)( \partial \bm{A} / \partial t)$.

\begin{table*}[h!tbp]
\caption{
Laser-induced damage thresholds (LIDT) of TiO$_2$ thin films and multilayers at different wavelengths and pulse durations (FWHM); these values are reported in previous works \cite{yao2008investigation,jupe2009calculations,sanz2009femtosecond,sanz2010ultra,negres201640,kumar2020laser}.
}
   \label{tab:lidt}
  \begin{ruledtabular}
    \begin{tabular}{ccccc}
      Photon energy (eV) & Wavelength (nm) & Pulse duration (fs) & Fluence (J/cm$^{2}$) & Reference \\
      \hline
      1.55 & 800 & 50  & 0.5 & Ref.~\onlinecite{yao2008investigation} \\
      1.55 & 800 & 220 & 0.6 & Ref.~\onlinecite{yao2008investigation} \\
      1.65--1.82 & 682 -- 750 & 130 & 0.4	& Ref.~\onlinecite{jupe2009calculations} \\
      1.82--2.10 & 590 -- 682 & 130 & 0.15 & Ref.~\onlinecite{jupe2009calculations} \\
      1.55 & 800 &	80	& 0.140	& Ref.~\onlinecite{sanz2009femtosecond} \\
      2.35 & 527 &	300	& 0.123	& Ref.~\onlinecite{sanz2010ultra} \\
      1.60 & 773 &	40 & 0.45	(TiO$_2$/$\mathrm{HfO}_2$) & Ref.~\onlinecite{negres201640} \\
      1.17 & 1064 &	$1 \times 10^4$ & 2.09 ($\mathrm{SiO}_2$/TiO$_2$)	& Ref.~\onlinecite{kumar2020laser}
    \end{tabular}
  \end{ruledtabular}
\end{table*}

\begin{figure}[h!tbp]
    \centering
    \includegraphics[width=\textwidth]{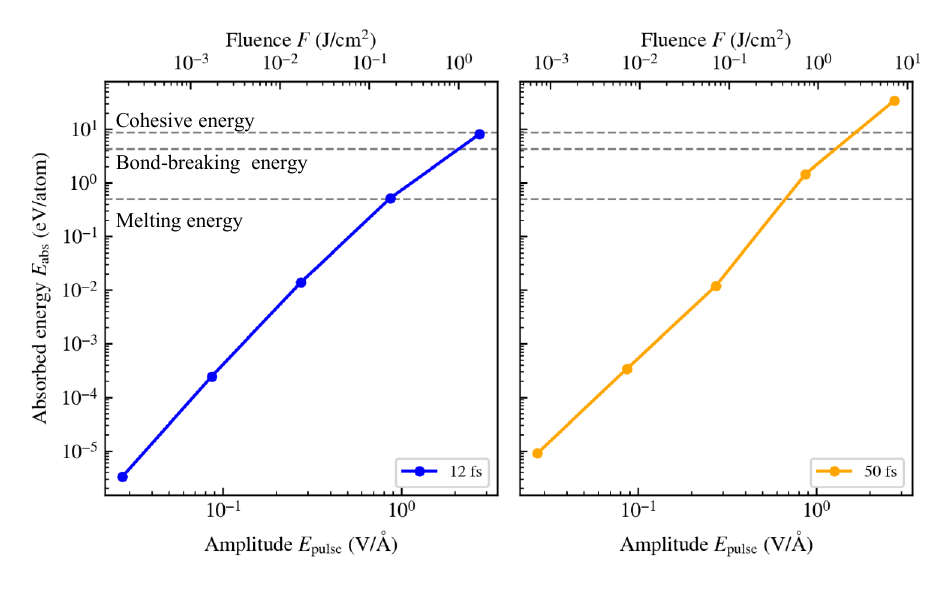}
    \caption{
      Dependence of the absorbed energy $\mathscr{E}_{\mathrm{abs}}$ on the pulse amplitude $E_\mathrm{pulse}$ for bulk rutile crystals.
      The dotted lines represent the melting energy (Eq.~\eqref{eq:e_melt}), cohesive energy (Eq.~\eqref{eq:e_coh}), and bond-breaking energy (Eq.~\eqref{eq:e_bond}).
      The corresponding pulse fluence $F$ is estimated as described in Appendix~\ref{sec:fluence_est}.
    }
    \label{fig:energy}
\end{figure}
Figure~\ref{fig:energy} shows the relationship between the pulse amplitude $E_\mathrm{pulse}$ and the resulting absorbed energy $\mathscr{E}_{\mathrm{abs}}$ for bulk rutile TiO$_2$.

In the high-intensity regime, nonlinear absorption processes, including multiphoton and free-carrier absorption, become significant; laser-induced damage resulting from these mechanisms can degrade the functionality of optical components, such as dielectric mirrors.
To estimate the ablation threshold through first-principles calculations, we compare the absorbed energy $\mathscr{E}_{\mathrm{abs}}$ with material-specific energetic thresholds \cite{sato2015time, venkat2022three}, specifically evaluating the melting energy $\mathscr{E}_{\mathrm{melt}}$, cohesive energy $\mathscr{E}_{\mathrm{coh}}$, and bond-breaking energy $\mathscr{E}_{\mathrm{bond}}$ as follows.

The melting energy, $\mathscr{E}_{\mathrm{melt}}$, is defined as the integral of the heat capacity from room temperature $\mathscr{T}_{\mathrm{room}}$
to the melting point $\mathscr{T}_{\mathrm{melt}}$:
\begin{align}
  \mathscr{E}_{\mathrm{melt}}
  &= \int_{\mathscr{T}_{\mathrm{room}}}^{\mathscr{T}_{\mathrm{melt}}} C_{p}(\mathscr{T}) \; \mathrm{d}\mathscr{T}
  \label{eq:e_melt}
  \;.
\end{align}
Here, $C_{p}(\mathscr{T})$ denotes the heat capacity at constant pressure as a function of the temperature $\mathscr{T}$ (see Appendix~\ref{sec:specific_heat}).
Based on experimental data \cite{chase1998nist}, $\mathscr{E}_{\mathrm{melt}}$ is estimated to be approximately $0.5~\mathrm{eV/atom}$.

The cohesive energy $\mathscr{E}_{\mathrm{coh}}$ is expressed as:
\begin{align}
  \mathscr{E}_{\mathrm{coh}}
  &= -\frac{
    \mathscr{E}_{\mathrm{bulk}} - N_{\mathrm{Ti}}\mu_{\mathrm{Ti}} - N_{\mathrm{O}}\mu_{\mathrm{O}}
  }{N_\mathrm{Ti} + N_\mathrm{O}}
  \;,
  \label{eq:e_coh}
\end{align}
where $\mathscr{E}_{\mathrm{bulk}}$ is the total energy, $N_{\mathrm{Ti}}$ and $N_{\mathrm{O}}$ are the numbers of Ti and O atoms, $\mu_{\mathrm{Ti}}$ and $\mu_{\mathrm{O}}$ represent the energies of isolated Ti and O atoms, respectively.
$\mathscr{E}_{\mathrm{coh}}$ can be estimated as $8.6~\mathrm{eV/atom}$.

Previous studies indicate that the ablation threshold typically lies between the melting and cohesive energies.
Empirically, the bond-breaking energy criterion \cite{venkat2022three} is expressed as follows:
\begin{align}
  \mathscr{E}_{\mathrm{bond}}
  &= \frac{\mathscr{E}_{\mathrm{coh}}}{n_{\mathrm{bond}}}
  \label{eq:e_bond}
  \;,
\end{align}
where
$n_{\mathrm{bond}}$ denotes the number of bonds per atom, so that $\mathscr{E}_{\mathrm{bond}}$ provides an estimate of the average bond-breaking energy, which can be estimated as $4.3~\mathrm{eV/atom}$.

As shown in Fig.~\ref{fig:energy}, the absorbed energy $\mathscr{E}_{\mathrm{abs}}$ increases rapidly with the incident field amplitude $E_\mathrm{pulse}$ and follows the power law $\mathscr{E}_{\mathrm{abs}} \propto E_\mathrm{pulse}^4$, consistent with a dominant two-photon absorption process.
For convenience, we provide the corresponding pulse fluence $F$ which is estimated as described in Appendix~\ref{sec:fluence_est}.
The bond-breaking energy threshold (Eq.~\eqref{eq:e_bond}) is reached at $E_\mathrm{pulse} \simeq 2.10~\mathrm{V}/\text{\AA}$ ($F \simeq 1.01~\mathrm{J}/\mathrm{cm}^2$) for $\tau_\mathrm{pulse}=12$~fs and at $E_\mathrm{pulse} \simeq 1.27~\mathrm{V}/\text{\AA}$ ($F \simeq 1.58~\mathrm{J}/\mathrm{cm}^2$) for $\tau_\mathrm{pulse}=50$~fs.
It should be noted that this fluence estimate is valid only in the low-intensity regime, where the optical response is linear.
At high intensities, however, nonlinear optical effects modify the reflectance \cite{boyd2008nonlinear}, requiring the more precise multiscale analysis detailed in Sec.~\ref{sec:multiscale}.
The estimated damage-threshold fluences are slightly higher than the experimental LIDT values reported in Table~\ref{tab:lidt}, but remain of the same order of magnitude.

Additionally, $\mathscr{E}_{\mathrm{abs}}$ clearly depends on the polarization direction in the weak-field region, reflecting the anisotropy of the dielectric function (see Supplementary Material S.~1).
This dependence vanishes in the high-intensity regime because the broad excitation of higher-energy bands overrides the optical selection rules near the band edge that dominate at lower intensities.

\subsection{Optical Reflectance in Thin Films}
\label{sec:multiscale}

\begin{figure*}[h!tbp]
    \centering
    \includegraphics[width=1.0\textwidth]{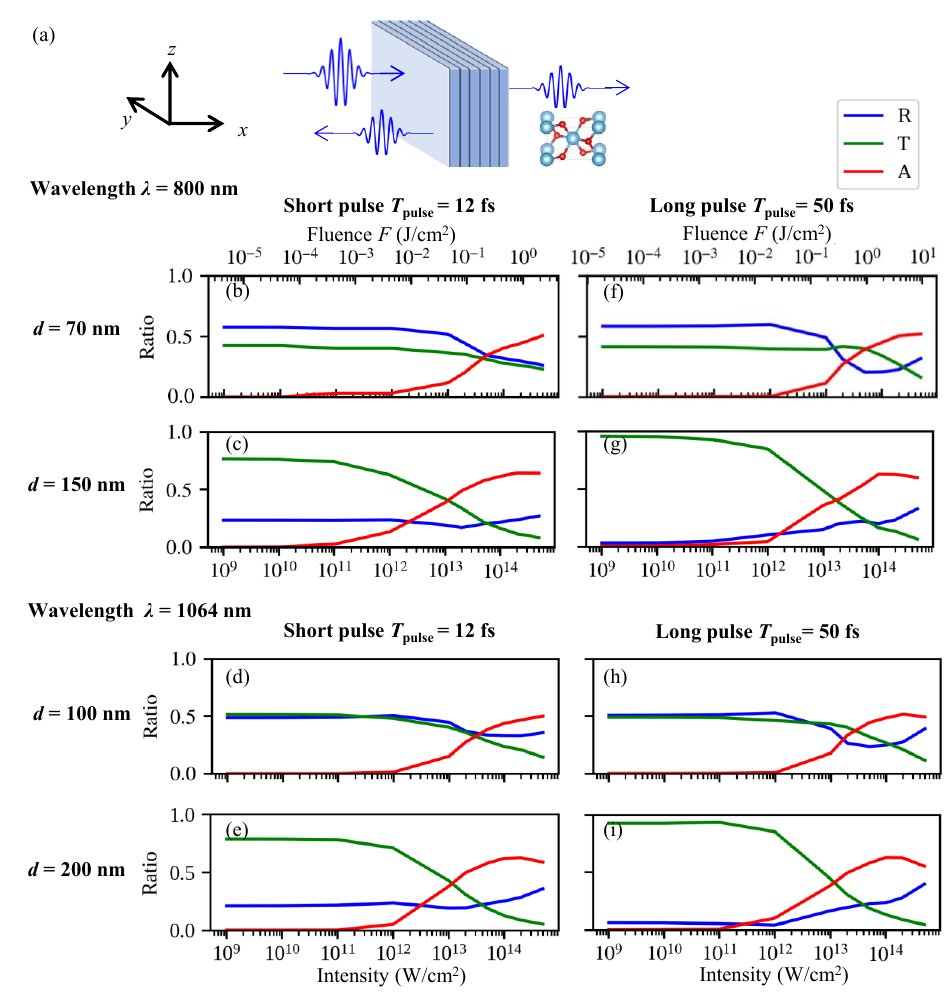}
    \caption{
      Calculated reflection ($R$), transmission ($T$), and absorption ($A$) coefficients of rutile TiO$_2$ thin films under ultrashort laser pulses.
      (a)~Schematic illustration of the geometry. (b)--(i) Dependence of reflectance, transmittance, and absorptance on intensity for pulse durations of $12~\mathrm{fs}$ and $50~\mathrm{fs}$, wavelengths of $800$~nm and $1064$~nm, and film thicknesses satisfying quarter-wave ($d \approx \lambda/4n_{\mathrm{eff}}$) and half-wave ($d \approx \lambda/2n_{\mathrm{eff}}$) conditions.
    }
    \label{fig:reflection}
\end{figure*}

Here, we predict the reflection and propagation processes of laser pulses through rutile thin films using the multiscale method.
As shown in Fig.~\ref{fig:reflection}(a), we consider a thin film with a thickness $d = 70$---$200$~nm, where the surface is defined by the (100) plane of rutile.
The incident laser pulses propagate along the $x$ axis and are linearly polarized along the $z$ axis.
The pulse intensity is varied from $10^9~\mathrm{W}/\mathrm{cm}^2$ to $5 \times 10^{14}~\mathrm{W}/\mathrm{cm}^2$ for two durations of $12~\mathrm{fs}$ and $50~\mathrm{fs}$.
We also calculate the reflectance $R$, transmittance $T$, and absorptance $A$ from incident, reflected, and transmitted wave (see Appendix~\ref{sec:calc_rta}).
The dependence of these quantities on the incident intensity $I$ is plotted in Fig.~\ref{fig:reflection}(b)--(i).
(For convenience, the fluence $F$ of the incident pulse, calculated using Eq.~\eqref{eq:fluence0}, is also indicated along the upper horizontal axes.)

For $(d,\lambda) = (70~\mathrm{nm}, 800~\mathrm{nm})$ (Fig.~\ref{fig:reflection}(b)), $A$ remains near zero and $R$ converges to a constant value in the low-intensity limit. As the incident intensity increases, the reflectance gradually decreases while the absorptance rises steadily. At higher intensities, the rapid increase in absorption is dominated by multiphoton excitation processes, as illustrated in Fig.~\ref{fig:energy}. Beyond this regime, a substantial population of optically excited free carriers yields a metallic-type response that strongly suppresses the reflectance. These qualitative features persist across different thickness and wavelength combinations (Figs.~\ref{fig:reflection}(c)--(e)).

We now examine the role of pulse duration. For $\tau_\mathrm{pulse}=12$~fs (Fig.~\ref{fig:reflection}(b)), $R$ falls to 90\% of its linear value at $I_{0.9} = 9.9 \times 10^{12}~\mathrm{W}/\mathrm{cm}^2$ and to 50\% at $I_{0.5} = 2.6 \times 10^{14}~\mathrm{W}/\mathrm{cm}^2$. A qualitatively similar trend is observed for the longer pulse of $\tau_\mathrm{pulse}=50$~fs (Fig.~\ref{fig:reflection}(f)), with $I_{0.9}=4.7 \times 10^{12}~\mathrm{W}/\mathrm{cm}^2$ and $I_{0.5}=2.4 \times 10^{13}~\mathrm{W}/\mathrm{cm}^2$.

In addition, for long pulse durations, the excitation can be approximated as continuous-wave illumination, and in the linear regime the reflectance is well described by standard thin-film Fresnel theory \cite{yeh1988optical,macleod2010thinfilm} (see Appendix~\ref{sec:thinfilm} for the derivation and analytical solution).
The analytical solution predicts maximum and minimum reflectance at quarter-wave and half-wave film thicknesses, $d = \lambda/4n_\mathrm{eff}$ and $d = \lambda/2n_\mathrm{eff}$, respectively.
At $\lambda=800$~nm, this corresponds to a maximum near $d\simeq 70$~nm (Fig.~\ref{fig:reflection}(f)) and a minimum near $d\simeq 150$~nm (Fig.~\ref{fig:reflection}(g)).
Similarly, at $\lambda = 1064$~nm, these occur at $d = 100$~nm (maximum, Fig.~\ref{fig:reflection}(h)) and $d = 200$~nm (minimum, Fig.~\ref{fig:reflection}(i)).
The simulated reflectance in the low-intensity limit agrees quantitatively with these predictions for each chosen thickness and wavelength. As observed for short pulses, increasing the incident intensity beyond the linear regime leads to a rapid increase in absorption accompanied by strong suppression of the reflectance.

\subsection{Dependence on Crystal Structure}

\begin{figure*}[h!tbp]
    \centering
    \includegraphics[width=1.0\textwidth]{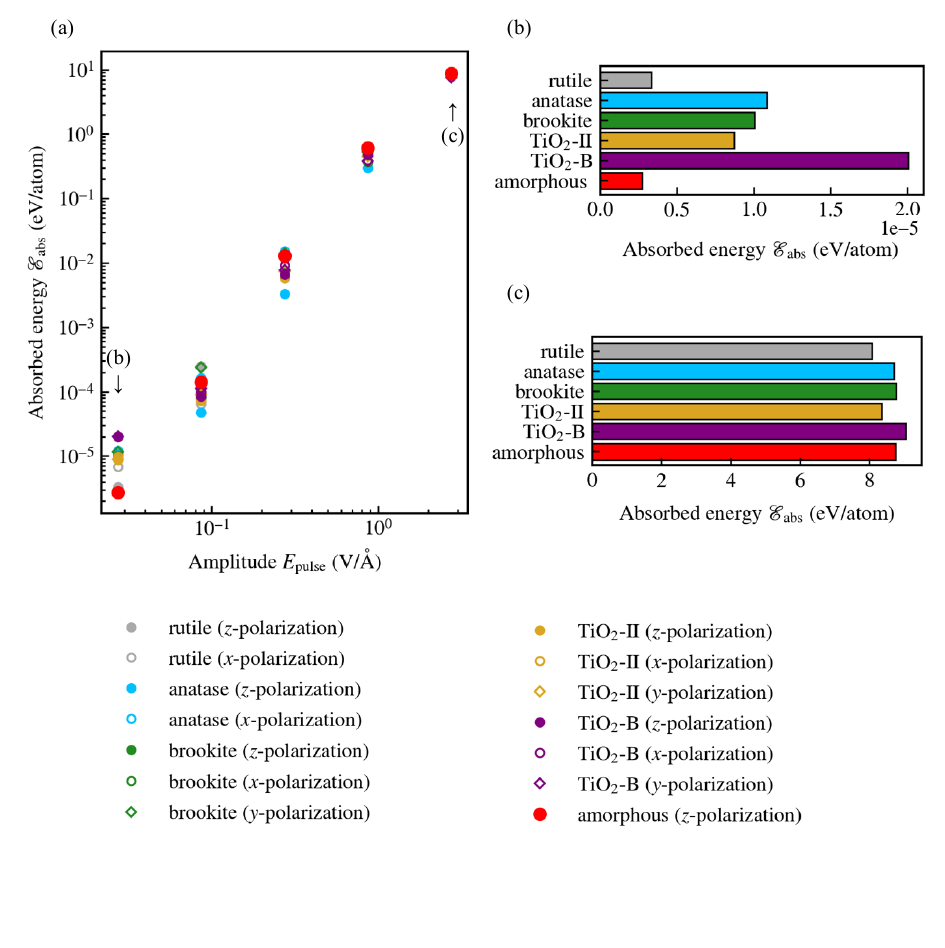}
\caption{
          Calculated absorbed energies $\mathscr{E}_{\mathrm{abs}}$ for various TiO$_2$ structures and incident light polarizations: (a)~$\mathscr{E}_{\mathrm{abs}}$ as a function of the applied field amplitudes, where the colors of the markers denote different models and polarization directions; (b)~absorbed energies under a weak incident electric field ($E=2.7 \times 10^{-2}~\mathrm{V}/\text{\AA}$); and (c)~absorbed energies under a strong incident electric field ($E=2.7~\mathrm{V}/\text{\AA}$). The details of the models are provided in Appendices~\ref{sec:param} and \ref{sec:amorphous}.
        }
    \label{fig:phase}
\end{figure*}

While the previous analyses focused on the rutile phase, industrially fabricated films often exhibit a mixture of crystalline phases or are entirely amorphous. To provide a realistic description of these materials and to investigate the influence of structural polymorphism on nonlinear optical interactions, we extend our comparative analysis to various stable and metastable crystalline polymorphs (rutile, anatase, brookite, TiO$_2$-B, and TiO$_2$-II) as well as an amorphous supercell model. The representative atomic structures for these phases are illustrated in Fig.~\ref{fig:model}.

Figure~\ref{fig:phase}(a) illustrates the dependence of the absorbed energy $\mathscr{E}_{\mathrm{abs}}$ on the incident electric field amplitude for various TiO$_2$ structures and polarizations.
Figures~\ref{fig:phase}(b) and (c) provide a detailed comparison of $\mathscr{E}_{\mathrm{abs}}$ at a weak field ($E=2.7 \times 10^{-2}~\mathrm{V}/\text{\AA}$) and a strong field ($E=2.7~\mathrm{V}/\text{\AA}$), respectively.

As shown in Fig.~\ref{fig:phase}(a), all crystalline phases and the amorphous structure exhibit the same intensity dependence of the absorbed energy as that in Fig.~\ref{fig:energy}, reflecting the multiphoton absorption process.
In the low-intensity regime, the absorbed energy exhibits variation depending on the crystal structure.
Specifically, $\mathscr{E}_{\mathrm{abs}}$ is relatively small in the rutile phase and the amorphous structure, whereas it is large in the TiO$_2$-II and TiO$_2$-B phases (Fig.~\ref{fig:phase}(b)); this tendency is independent of the polarization direction of the incident light.
Furthermore, as the incident light intensity increases, the structural dependence of the absorbed energy decreases.
In the high-intensity regime where material damage can occur, all structures exhibit similar energy absorption characteristics (Fig.~\ref{fig:phase}(c)).
In a related context, Yamada and Yabana reported on the reflectance of oxide dielectric crystals such as $\mathrm{SiO}_2$ in a previous study \cite{yamada2024interaction}, showing that the optical response is dominated by plasma reflection at extremely high incident intensities.
In addition, these observations suggest that bulk crystalline TiO$_2$ may serve as a reasonable computational surrogate for industrially produced amorphous thin films, provided that the analysis is restricted to the high-intensity regime.

\section{Conclusion}
In this study, we investigated the nonlinear optical response of TiO$_2$ under intense ultrashort laser pulses using first-principles simulations.
In the case of the bulk material, the optical response transitions from linear at low intensities to a regime dominated by two-photon absorption as the intensity increases.
Furthermore, by employing a multiscale Maxwell--TDDFT method, we simulated the propagation and reflection of laser pulses through thin films while accounting for light--matter interactions.
Our results demonstrate that in the low-intensity regime, TiO$_2$ exhibits a linear response consistent with Fresnel reflection.
However, as the incident intensity increases, evaluation of the reflected and transmitted fields reveals reduced reflectance at high intensities ($I \sim 10^{13}~\mathrm{W}/\mathrm{cm}^2$), demonstrating a pronounced nonlinear optical response.

Moreover, we provided a first-principles estimate of the ablation threshold by comparing the absorbed energy with critical energetic thresholds.
The predicted damage threshold is of the same order of magnitude as the experimental values, indicating that our approach captures the characteristic energy scale of material degradation from a microscopic perspective.

These findings provide critical microscopic insights into the optical degradation mechanisms of TiO$_2$-based dielectric components. Such insights are essential for advancing toward new frontiers in high-intensity physics, such as observing the nonlinear QED effects, and enabling extreme engineering applications including space debris removal.
Both pursuits require reflective components that can withstand extreme field strengths without degradation; therefore, robust dielectric multilayers are desirable.
Consequently, the theoretical framework established in this work not only elucidates the limits of current materials but also provides a predictive tool for the design and optimization of high-power dielectric mirrors.

\section*{Supplementary Material}
The supplementary material provides additional computational results, including the anisotropy of the dielectric function in bulk rutile TiO$_2$ (S.~1) and the structure and dielectric function of amorphous TiO$_2$ (S.~2).

\begin{acknowledgments}
The authors are grateful to Dr.~Shunsuke Yamada and Dr.~Tomohiro Otobe of the National Institutes for Quantum Science and Technology (QST), as well as Prof.~Tomoya Ono of Kobe University, for helpful discussions.
This work was partially supported by MEXT through JSPS KAKENHI Grants (Nos.~24K01224 and 24K06922).
Furthermore, this project received support from the Collaborative Research Program of the Institute of Laser Engineering (ILE) at the University of Osaka (No.~2026B2-047).
Numerical calculations were performed using the computational facilities of the Institute for Solid State Physics (ISSP) at the University of Tokyo, the supercomputer Miyabi system provided by the Multidisciplinary Cooperative Research Program (MCRP) at the Center for Computational Sciences, University of Tsukuba, and the supercomputer Fugaku provided by the RIKEN Center for Computational Science (hp230175, hp240178, and hp250193).
ChatGPT (OpenAI) and Claude (Anthropic) were used for English translation and grammatical checking. The authors reviewed and revised all AI-generated text and take full responsibility for the final manuscript.
\end{acknowledgments}

\section*{Author Declarations}

\subsection*{Conflict of Interest}
The authors have no conflicts to disclose.

\subsection*{Author Contributions}
\noindent \textbf{Koya Shimaoka:} Methodology (equal); Software (lead); Validation (equal); Formal analysis (lead); Investigation (lead); Data curation (lead); Writing--original draft (equal); Visualization (equal).\par
\noindent \textbf{Yusuke Kondo:} Conceptualization (equal); Writing--review \& editing (equal); Visualization (equal).\par
\noindent \textbf{Kazunori Shibata:} Conceptualization (equal); Writing--review \& editing (equal); Project administration (lead).\par
\noindent \textbf{Mitsuharu Uemoto:} Conceptualization (equal); Methodology (equal); Validation (equal); Investigation (supporting); Writing--original draft (equal); Writing--review \& editing (equal); Supervision (lead); Funding acquisition (lead).\par

\section*{Data Availability}
The data that support the findings of this study are available from the corresponding author upon reasonable request.

\bibliographystyle{apsrev4-2}
\bibliography{refs}

\appendix

\section{Pulse Shape}
\label{sec:pulse_shape}
In our calculations, the pump pulse applied to the material is modeled using a $\sin^2$ envelope, which is widely employed in previous studies \cite{noda2019salmon,yamada2024interaction,uemoto2021first,li2026three}.
The corresponding vector potential is given by
\begin{align}
\bm{A}_\mathrm{pulse}(t)
=&
-\frac{c E_\mathrm{pulse}}{\omega_\mathrm{pulse}}
\sin\left[
\omega_\mathrm{pulse}\left(t-\frac{\tau_\mathrm{pulse}}{2}\right)
\right]
\sin^2\left(\frac{\pi t}{\tau_\mathrm{pulse}}\right)
\hat{\bm{e}}_\mathrm{pulse},
\end{align}
where $E_\mathrm{pulse}$, $\omega_\mathrm{pulse}$, and $\hat{\bm{e}}_\mathrm{pulse}$ denote the electric-field amplitude, the central angular frequency, and the polarization vector, respectively, and $\tau_\mathrm{pulse}$ is the pulse duration. The pulse duration is related to the full width at half maximum (FWHM) by $T_\mathrm{FWHM} \simeq 0.36\,\tau_\mathrm{pulse}$.

\section{Estimation of Pulse Fluence}
\label{sec:fluence_est}
The pulse fluence in vacuum is generally defined as
\begin{align}
F_\mathrm{vacuum} =&
\frac{c}{4\pi}
\int_0^\infty
\left[E(t)\right]^2 \;
\mathrm{d}t
\;,
\label{eq:fluence0}
\end{align}
where $E(t)$ represents the time-dependent electric field.
For a uniform solid medium with refractive index $n_\mathrm{eff}$, Eq.~\eqref{eq:fluence0} can be adapted to yield
\begin{align}
F_\mathrm{solid} =&
\frac{c n_\mathrm{eff}}{4\pi}
\int_0^\infty
\left[E(t)\right]^2 \;
\mathrm{d}t
\;.
\label{eq:fluence_media}
\end{align}

A laser pulse incident normally on a solid medium from vacuum is partially reflected at the surface.
The incident fluence $F_\mathrm{laser}$ can be estimated from the fluence inside the solid as
\begin{align}
F_\mathrm{laser}
=&
\frac{F_\mathrm{solid}}{1 - |r|^2}
\;,
\label{eq:fluence_laser}
\end{align}
where $r$ is the Fresnel reflection coefficient for normal incidence \cite{hecht2015optics}.
Here, $r$ is given by
\begin{align}
r
=& \frac{n_\mathrm{eff}-1}{n_\mathrm{eff}+1}
\;.
\label{eq:fresnel_r}
\end{align}
The factor $(1-|r|^2)$ in Eq.~\eqref{eq:fluence_laser} represents the fraction of the incident pulse energy transmitted into the material. The incident fluence $F_\mathrm{laser}$ is denoted by $F$ in Sec.~\ref{sec:pulse}.
At $\hbar\omega=1.55~\mathrm{eV}$, the effective refractive index is estimated from Fig.~\ref{fig:epsilon} as $n_\mathrm{eff}=\sqrt{\epsilon}\simeq 2.8$.
This estimate assumes that the electric field is weak to justify the linear-response approximation.

\section{Estimation of Reflectance, Transmittance, and Absorptance in Thin Films}
\label{sec:calc_rta}
The reflectance $R$ and transmittance $T$ of a thin film irradiated by a laser pulse are defined as
\begin{align}
R &= \frac{F_\mathrm{ref}}{F_\mathrm{inc}}
\;,
\label{eq:reflectance}
\\
T &= \frac{F_\mathrm{tra}}{F_\mathrm{inc}}
\;,
\label{eq:transmittance}
\end{align}
where $F_\mathrm{inc}$ is the fluence of the incident pulse in vacuum, evaluated using Eq.~\eqref{eq:fluence0}.
Similarly, $F_\mathrm{ref}$ and $F_\mathrm{tra}$ denote the fluences of the waves reflected and transmitted from the thin film, respectively.
By energy conservation, the absorptance $A$ is given by
\begin{align}
A &= 1 - R - T
\;.
\label{eq:absorptance}
\end{align}

\section{Specific Heat of TiO$_2$}
\label{sec:specific_heat}
The heat capacity $C_p$ of rutile phase TiO$_2$ can be approximated by the following empirical function \cite{chase1998nist}:
\begin{align}
  C_p &= A + B t + C t^2 + D t^3 + \frac{E}{t^2}
\end{align}
with
  $A = 67.29830$,
  $B = 18.70940$,
  $C = -11.57900$,
  $D = 2.449561$,
  $E = -1.485471$, and
  $t = \mathscr{T} / 1000$,
where $C_p$ is expressed in $\mathrm{J\,mol^{-1}\,K^{-1}}$ and $\mathscr{T}$ denotes the temperature in $\mathrm{K}$.

\section{Normal-Incidence Optical Response of an Anisotropic Slab}
\label{sec:thinfilm}

We consider an air/anisotropic-slab/air structure at normal incidence. The slab is taken to be linear, nonmagnetic, and lossless. Its relative dielectric tensor is written in diagonal form,
\begin{equation}
\bm{\varepsilon}
=
\operatorname{diag}(\varepsilon_x,\varepsilon_y,\varepsilon_z).
\label{eq:eps_tensor}
\end{equation}
The slab lies in the $y$--$z$ plane, and its surface normal is parallel to $x$. To distinguish the propagation coordinate from the crystallographic $z$ axis and the $E\parallel z$ polarization, we introduce $s$ as the coordinate along the surface normal, increasing from the incident medium toward the exit medium. The slab thus occupies $0\le s\le d$.

At normal incidence, the two tangential principal-axis polarizations, $E\parallel y$ and $E\parallel z$, are uncoupled eigenpolarizations. For $j\in\{y,z\}$, the corresponding effective refractive index is
\begin{equation}
 n_{\mathrm{eff},j}=\sqrt{\varepsilon_j}.
\label{eq:neff}
\end{equation}
Accordingly, the present geometry reduces exactly to two independent scalar thin-film problems. More general propagation directions or rotated principal axes require the full anisotropic transfer-matrix formulation \cite{yeh1988optical,berreman1972optics}.

For either eigenpolarization, the electric field normalized by the incident amplitude is written piecewise as
\begin{equation}
\frac{E_j(s)}{E_{\mathrm{inc}}}
=
\begin{cases}
 e^{ik_{\mathrm{in}}s}+r_j e^{-ik_{\mathrm{in}}s}, & s<0,\\[3pt]
 C_j e^{ik_js}+D_j e^{-ik_js}, & 0\le s\le d,\\[3pt]
 t_j e^{ik_{\mathrm{out}}(s-d)}, & s>d,
\end{cases}
\label{eq:piecewise_field}
\end{equation}
where
\begin{equation}
 k_0=\frac{2\pi}{\lambda_0},
 \qquad
 k_{\mathrm{in}}=n_0k_0,
 \qquad
 k_j=n_{\mathrm{eff},j}k_0,
 \qquad
 k_{\mathrm{out}}=n_{\mathrm{out}}k_0.
\label{eq:wavenumbers}
\end{equation}
Here, $r_j$ and $t_j$ are the complex reflection and transmission amplitudes, whereas $C_j$ and $D_j$ are the forward- and backward-propagating amplitudes inside the slab.

For nonmagnetic media, continuity of the tangential electric and magnetic fields at $s=0$ and $s=d$ gives the four independent conditions
\begin{subequations}
\label{eq:boundary_conditions}
\begin{gather}
 1+r_j=C_j+D_j,
 \label{eq:bc_front_e}
 \\
 n_0(1-r_j)=n_{\mathrm{eff},j}(C_j-D_j),
 \label{eq:bc_front_h}
 \\
 C_j e^{ik_jd}+D_j e^{-ik_jd}=t_j,
 \label{eq:bc_back_e}
 \\
 n_{\mathrm{eff},j}
 \left(C_j e^{ik_jd}-D_j e^{-ik_jd}\right)
 =n_{\mathrm{out}}t_j.
 \label{eq:bc_back_h}
\end{gather}
\end{subequations}
For the air/slab/air configuration considered here, $n_0=n_{\mathrm{out}}=1$. The power reflectance, transmittance, and absorptance are then \cite{yeh1988optical,macleod2010thinfilm}
\begin{equation}
 R_j=|r_j|^2,
 \qquad
 T_j=\frac{n_{\mathrm{out}}}{n_0}|t_j|^2,
 \qquad
 A_j=1-R_j-T_j.
\label{eq:rta}
\end{equation}
Because all refractive indices are real, the model is lossless and $A_j=0$ apart from numerical round-off error. The quarter-wave and half-wave physical thicknesses are defined by
\begin{equation}
 d_{\lambda/4,j}=\frac{\lambda_0}{4n_{\mathrm{eff},j}},
 \qquad
 d_{\lambda/2,j}=\frac{\lambda_0}{2n_{\mathrm{eff},j}}.
\label{eq:optical_thicknesses}
\end{equation}

Table~\ref{tab:summary} separately summarizes (i) the optical coefficients at $d=\SI{100}{nm}$ and (ii) the corresponding quarter-wave and half-wave physical thicknesses for each wavelength and eigenpolarization. At \SI{800}{nm}, $T$ and $R$ are both close to 0.5 for the two eigenpolarizations. At \SI{1064}{nm}, the reflectance is higher, especially for $E\parallel z$, for which $R=0.6182$. The absorptance remains essentially zero because only real refractive indices are used. The quarter-wave thicknesses are 72.12 and 66.90~nm at \SI{800}{nm}, and 98.59 and 91.18~nm at \SI{1064}{nm}, for $E\parallel y$ and $E\parallel z$, respectively.

\begin{table}[h!tbp]
\centering
\caption{Calculated optical coefficients at $d=\SI{100}{nm}$ and the corresponding quarter-wave and half-wave thicknesses for the two eigenpolarizations.}
\label{tab:summary}
\small
\renewcommand{\arraystretch}{1.16}

(a) Optical response \par\vspace{3pt}
\setlength{\tabcolsep}{8.0pt}
\begin{ruledtabular}
\begin{tabular}{cccccc}
$\lambda_0$ (nm) & Polarization & $n_{\mathrm{eff}}$ & $T$ & $R$ & $A$ \\
\hline
800  & $E\parallel y$ & 2.7732 & 0.5047 & 0.4953 & $\simeq 0$ \\
800  & $E\parallel z$ & 2.9894 & 0.5275 & 0.4725 & $\simeq 0$ \\
1064 & $E\parallel y$ & 2.6980 & 0.4249 & 0.5751 & $\simeq 0$ \\
1064 & $E\parallel z$ & 2.9174 & 0.3818 & 0.6182 & $\simeq 0$
\end{tabular}
\end{ruledtabular}

\vspace{8pt}
(b) Characteristic physical thicknesses\par\vspace{3pt}
\setlength{\tabcolsep}{11.0pt}
\begin{ruledtabular}
\begin{tabular}{cccc}
$\lambda_0$ (nm) & Polarization & \makecell{Quarter-wave
thickness (nm) \\$d_{\lambda/4,j}=\lambda_0/(4n_{\mathrm{eff},j})$} & \makecell{Half-wave thickness (nm) \\$d_{\lambda/2,j}=\lambda_0/(2n_{\mathrm{eff},j})$} \\
\hline
800  & $E\parallel y$ & 72.12 & 144.24 \\
800  & $E\parallel z$ & 66.90 & 133.81 \\
1064 & $E\parallel y$ & 98.59 & 197.18 \\
1064 & $E\parallel z$ & 91.18 & 182.35
\end{tabular}
\end{ruledtabular}
\end{table}

\begin{figure}[h!tbp]
\centering
\includegraphics[width=0.98\linewidth]{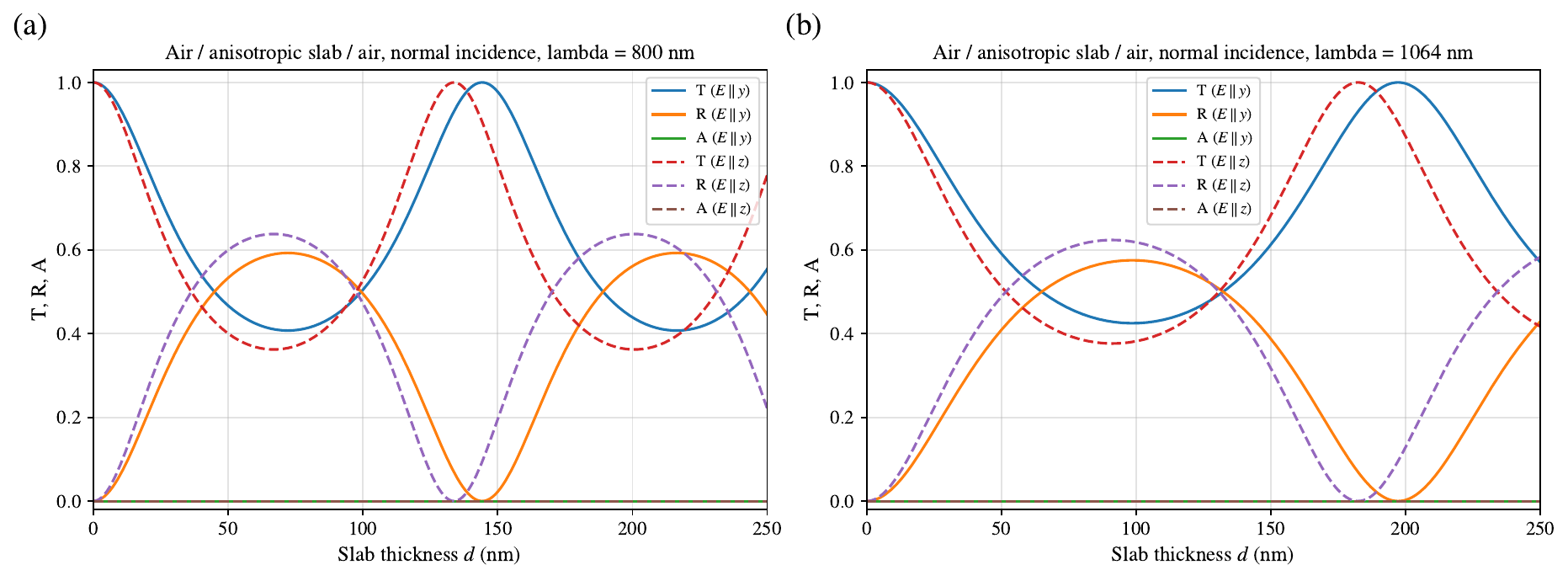}
\caption{Thickness dependence of the transmittance, reflectance, and absorptance at normal incidence for (a) $\lambda_0=\SI{800}{nm}$ and (b) $\lambda_0=\SI{1064}{nm}$. Solid and dashed curves denote the $E\parallel y$ and $E\parallel z$ eigenpolarizations, respectively. Real refractive indices imply $A=0$ and $R+T=1$.}
\label{fig:rta}
\end{figure}

The thickness-dependent optical coefficients are shown in Fig.~\ref{fig:rta}. The periodic exchange between $T$ and $R$ is the usual Fabry--P\'erot response of a lossless film. The small polarization-dependent shift of the extrema follows directly from $n_{\mathrm{eff},y}\ne n_{\mathrm{eff},z}$.

\section{Crystal Structures and Computational Parameters}
\label{sec:param}
In this study, we construct computational models for several crystalline phases of TiO$_2$.
To optimize the efficiency of the real-space representation, rectangular supercells are adopted for most phases instead of conventional primitive cells (see Fig.~\ref{fig:model}).
The spatial grid spacing is consistently adjusted to match that of the rutile phase, ensuring a uniform level of numerical precision across different polymorphs.
Detailed structural and computational parameters are summarized in Table~\ref{tab:cell_params}.
\begin{table}[t]
  \caption{Computational parameters for the various crystalline phases of TiO$_2$.}
  \label{tab:cell_params}
  \begin{ruledtabular}
    \begin{tabular}{cccc}
    Phase & Cell size ($\text{\AA} \times
    ~\text{\AA} \times ~\text{\AA}$) & Realspace grid & $k$-point grid \\
    \hline
    Rutile &
    $4.5~\times 4.5~\times 2.8$ &
    $24 \times 24 \times 16$ &
    $4 \times 4 \times 6$
    \\
    Anatase &
    $3.8~\times 3.8~\times 9.6$ &
    $24 \times 24 \times 60$ &
    $8 \times 8 \times 3$
    \\
    Brookite &
    $5.2~\times 5.5~\times 9.2$ &
    $36 \times 38 \times 64$ &
    $4 \times 4 \times 2$
    \\
    TiO$_2$-B &
    $ 12.3~\times 3.8~\times 6.6$ &
   $58 \times 18 \times 30$ &
   $2 \times 6 \times 4$
   \\
   TiO$_2$-II &
   $ 4.6~\times 4.9~\times 5.5$ &
   $28 \times 30 \times 34$ &
   $6 \times 6 \times 5$
    \end{tabular}
  \end{ruledtabular}
\end{table}

\section{Construction of Amorphous Models Using Machine-Learning Molecular Dynamics}
\label{sec:amorphous}
For the first-principles analysis of realistic amorphous TiO$_2$, we construct a large supercell model incorporating atomic-scale distortions. Machine-learning molecular dynamics (MLMD) simulations are performed to generate these atomic structure models via a melt-and-quench scheme. Initially, a rutile supercell containing 24 Ti and 48 O atoms is prepared by scaling the original rutile unit cell shown in Fig.~\ref{fig:model}(a) by $2 \times 2 \times 3$. Although the nominal supercell dimensions are $9.0~\text{\AA} \times 9.0~\text{\AA} \times 8.4~\text{\AA}$ based on the bulk lattice constants, we artificially transform the supercell into a cube of equal volume to ensure isotropy in the model.
By using finite-temperature MD, the initial structure is gradually heated from $0~\mathrm{K}$ to $5000~\mathrm{K}$ over $3~\mathrm{ps}$ and subsequently annealed at this temperature for $2~\mathrm{ps}$; the resulting liquid system is then quenched at a cooling rate of $100~\mathrm{K}/\mathrm{ps}$.

For these computations, we employ the MD methods implemented in the atomic simulation environment (ASE) library \cite{hjorth2017atomic} with the MACE-MPA-0 foundation model \cite{batatia2025foundation} as the interatomic machine-learning force field; this model is trained on the Materials Project dataset and accurately describes the structural stability of inorganic solid crystals.
The MD calculations are conducted under the NVT ensemble (constant number of atoms, volume, and temperature) with a time step of $1.0~\mathrm{fs}$.
In addition, we generate a variety of structural models with different initial conditions and computational parameters. We also verified that our amorphous structures exhibit sufficient disorder and isotropy by analyzing the radial distribution functions and the dielectric tensors (see Supplementary Material S.~2 for details).
\end{document}


\maketitle
\footnotetext[1]{
    Department of Electrical and Electronic Engineering,
    Graduate School of Engineering, Kobe University,
    Nada, Kobe 657-8501, Japan
}
\footnotetext[2]{
    Osaka Research Institute of Industrial Science and Technology (ORIST),
    2-7-1 Ayumino, Izumi, Osaka 594-1157, Japan
}
\footnotetext[3]{
    Institute of Laser Engineering (ILE), The University of Osaka,
    2-6 Yamadaoka, Suita, Osaka 565-0871, Japan
}

\renewcommand{\thesection}{S.\arabic{section}}
\numberwithin{figure}{section}
\numberwithin{equation}{section}
\numberwithin{table}{section}
\renewcommand{\thefigure}{\thesection.\arabic{figure}}
\renewcommand{\theequation}{\thesection.\arabic{equation}}
\renewcommand{\thetable}{\thesection.\arabic{table}}

\clearpage
\section{Anisotropy of the dielectric function in bulk rutile $\mathrm{TiO_2}$}

\begin{figure}[htbp]
    \centering
    \includegraphics[width=0.8\textwidth]{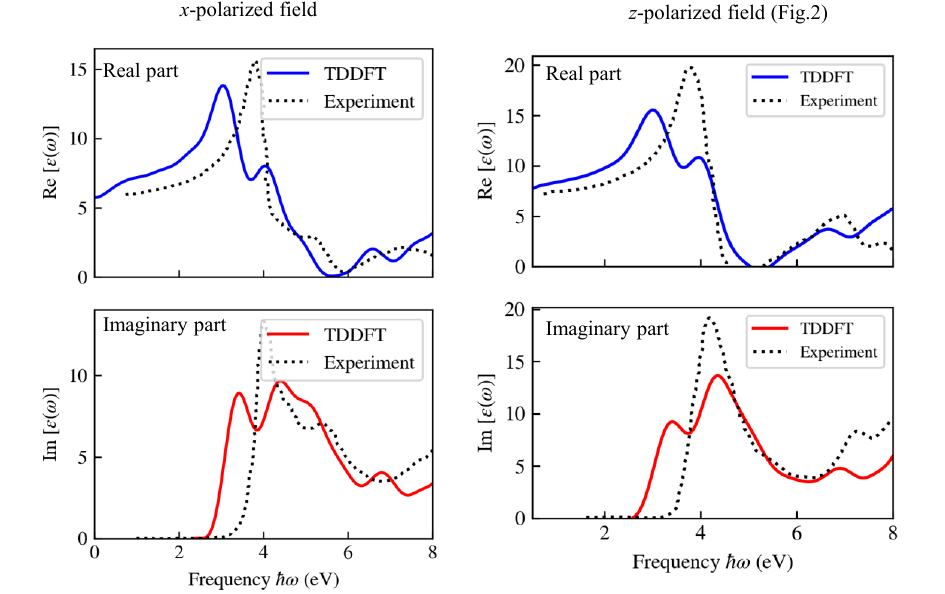}
    \caption{
        \label{fig:dielec_xz}
        Dielectric functions of bulk rutile $\text{TiO}_2$ for $x$- and $z$-polarization.
    }
\end{figure}

\clearpage
\section{Structure and dielectric function of amorphous $\text{TiO}_2$}

\begin{figure}[htbp]
    \centering
    \includegraphics[width=0.6\textwidth]{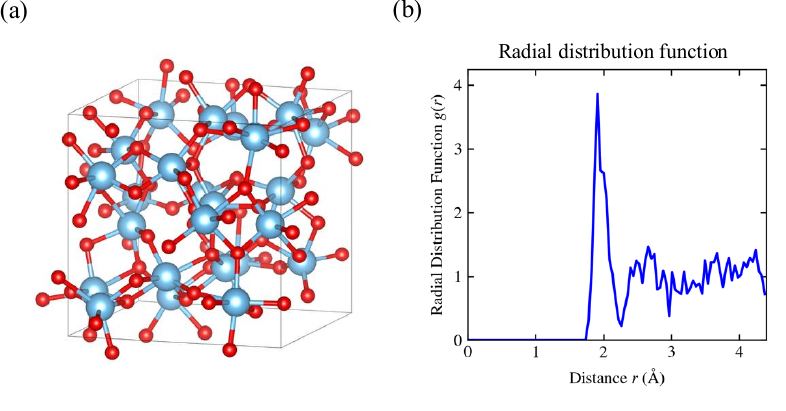}
    \caption{
        (a) Amorphous $\text{TiO}_2$ model and (b) radial distribution function.
        \label{fig:rdf}
    }
\end{figure}

\begin{figure}[htbp]
    \centering
    \includegraphics[width=1.0\textwidth]{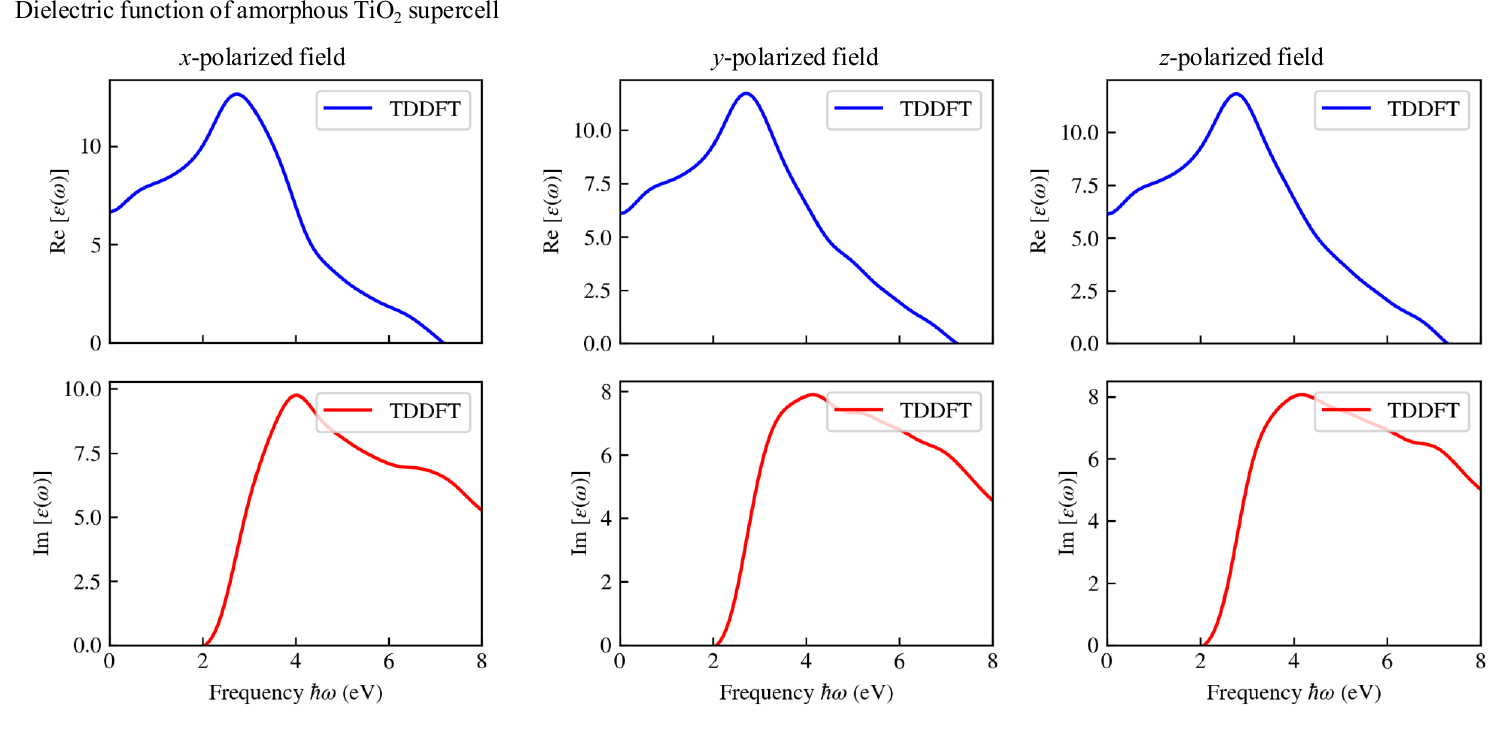}
    \caption{
        \label{fig:dielectric_amorphous}
        Components of the dielectric function for $x$-, $y$-, and $z$-polarization for the amorphous $\text{TiO}_2$ model.
    }
\end{figure}